\documentclass[11pt,a4paper,reqno]{amsart}
\usepackage{graphicx} %
\usepackage[left=2cm, right=2cm]{geometry}

\usepackage{times} %
\usepackage{amsmath} %
\usepackage{amssymb}  %
\usepackage{amsthm}
\usepackage{latexsym}
\usepackage{amsfonts,bbm}
\usepackage{xcolor}
\usepackage{mathtools}
\usepackage{enumerate}
\usepackage{cite}
\usepackage{tikz}
\usepackage{graphicx,caption}
\usepackage{todonotes}
\usepackage{float}
\usepackage{nicefrac}
\usepackage[foot]{amsaddr}
\usepackage{color}
\usepackage{graphicx}
\usepackage{bbm}

\usepackage{tikz}
\usetikzlibrary{shapes.geometric}
\usetikzlibrary{arrows.meta,arrows}

\usepackage{tabularx}
\usepackage{comment}
\usepackage[hidelinks, colorlinks=false]{hyperref}
\usepackage{cleveref}

\usepackage[normalem]{ulem}
\usepackage{algorithm}

\newtheorem{theorem}{Theorem}
\newtheorem{remark}[theorem]{Remark}
\newtheorem{proposition}[theorem]{Proposition}

\newtheorem{lemma}[theorem]{Lemma}
\newtheorem{assumption}[theorem]{Assumption}
\newtheorem{definition}[theorem]{Definition}

\newtheorem{problem}{Problem}

\newcommand{\rline}  {{\mathbb R}}
\newcommand{\xline}  {{\mathbb X}}

\newcommand{\nline}  {{\mathbb N}}
\newcommand{\dist}{\mathrm{dist}}

\newcommand{\kscr} {\mathcal{K}}
\newcommand{\xscr} {\mathbb{X}}
\newcommand{\uscr} {\mathbb{U}}

\newcommand{\mscr} {\mathcal{M}}
\newcommand{\sscr}{\mathcal{S}}
\newcommand{\cscr}{\mathcal{C}}
\newcommand{\aut}{\mathrm{aut}}

\newcommand{\bbm}[1]{\left[\begin{matrix} #1 \end{matrix}\right]}

\usepackage{amsmath,amssymb,amsfonts}
\usepackage{latexsym}
\usepackage{mathrsfs}
\usepackage{bbm}
\usepackage{xcolor}
\usepackage{tikz}
\usepackage{url}
\usepackage{stmaryrd}
\usepackage{dsfont}
\usepackage{placeins}
\usepackage{lipsum}
\usepackage{amsfonts}
\usepackage{graphicx}
\usepackage{epstopdf}
\usepackage{amsopn}
\ifpdf
  \DeclareGraphicsExtensions{.eps,.pdf,.png,.jpg}
\else
  \DeclareGraphicsExtensions{.eps}
\fi

\allowdisplaybreaks

\title{Energy-optimal control of discrete-time port-Hamiltonian systems}
\author{Arijit Sarkar$^{1}$}
\address{$^1$Department of Control Systems and Network Control Technology, Brandenburg University of Technology Cottbus-Senftenberg, Germany}
\author{Vaibhav Kumar Singh$^{2}$}
\address{$^{2}$Faculty of Science and Engineering, Jan C. Willems Center for Systems and Control, ENTEG, University of Groningen, The Netherlands}
\author{Manuel Schaller$^{3}$}
\address{$^{3}$Faculty of Mathematics, Chemnitz University of Technology, Germany}
\author{Karl Worthmann$^{4}$}
\address{$^{4}$Optimization-based Control Group, Institute of Mathematics, Technische Universität Ilmenau, Germany.}
\thanks{A.S.\ is supported by the Federal Ministry of Education and Research under project 03SF0693A, the Federal Government via the Coal Regions Investment Act (Investitionsgesetz Kohleregionen) under project 85056897 and co-financed with funds from the state of Brandenburg. V.K.S.\ is supported by the Dutch Research Council under Grant ESI.2019.005. M.S.\ and KW.\ gratefully acknowledge funding by the Deutsche Forschungsgemeinschaft (DFG, German Research Foundation) – Project-ID 519323897.}

\begin{document}

\begin{abstract}
   In this letter, we study the energy-optimal control of nonlinear port-Hamiltonian (pH) systems in discrete time. For continuous-time pH systems, energy-optimal control problems are strictly dissipative by design. This property, stating that the system to be optimized is dissipative with the cost functional as a supply rate, implies a stable long-term behavior of optimal solutions and enables stability results in predictive control.
   In this work, we show that the crucial property of strict dissipativity is not straightforwardly preserved by any energy-preserving integrator such as the implicit midpoint rule. Then, we prove that discretizations via difference and differential representations lead to strictly dissipative discrete-time optimal control problems. Consequently, we rigorously show a stable long-term behavior of optimal solutions in the form of a manifold (subspace) turnpike property. Finally, we validate our findings using two numerical examples.\smallskip
    

\end{abstract}
\maketitle

\section{Introduction}\label{sec:intro}
In recent years, port-Hamiltonian (pH) systems have drawn increasing attention across various fields of science and engineering as a highly structured framework for modular, energy-based modeling, analysis and control of dynamical systems, see \cite{JacZ12,SchaJelt14,MehU22}. By design, these models are passive, i.e., dissipative with a particular supply rate given by the supplied energy, enabling sophisticated techniques for passivity-based controller design and control by interconnection~\cite{Ortega02}.  While passivity-based control design usually relies on energy-shaping, the controller design in
optimization-based control, such as model predictive control (MPC; [5]), is based on solution of optimal control problems. This allows to naturally incorporate suitable performance criteria for the closed-loop via the cost functional. %
The respective design is crucial to ensure closed-loop stability and performance~\cite{MuelWort18}. There are also various inverse optimality results for nonlinear feedback controllers (such as damping injection), i.e., they construct an optimal control problem for which the feedback is optimal, see \cite[Section 3.5]{Sch17} or \cite[Section 3]{SepJanKok12}.
For pH systems, the supplied energy is readily available as a time integral of the product of input and output, yielding a very natural candidate for the cost functional to achieve a control task in an economic way, see \cite{ScZeWo:24} for an application to energy-optimal building stabilization. 
For {continuous-time} {linear} systems, it has been shown that the resulting optimal control problem is strictly dissipative, i.e., the dynamics are strictly dissipative with the cost functional as a supply rate, providing a stable long-term behavior of optimal solutions towards conservative subspaces~\cite{Schaller21}, see also~\cite{Faulwasser2022a} and \cite{philipp2024optimal} for extensions to differential-algebraic and nonlinear irreversible systems, respectively. Hence, in continuous-time systems, this crucial property serves as the foundation for performance and stability results in model predictive control~\cite{ZanoFaul18}.

However, while discretizations of the underlying optimal control problem are a central part of actual implementations in MPC, %
the impact of discretizations on the strict dissipativity in energy-optimal control of pH systems has only recently been explored in \cite{csen2023stage}.
In this work, we show that discretizations that ensure mere dissipativity of the discrete dynamics (such as, e.g., the implicit midpoint rule, see e.g.~\cite{kotyczka2019}) do not straightforwardly yield strict dissipativity of the optimal control problem. Thus, we suggest an advanced technique given by difference and differential representations (DDRs), which was proposed in~\cite{Monaco1998} and extended to discrete-time pH systems in~\cite{Moreschini2019}. 
Using this representation, strict dissipativity of the optimal control problem is always preserved for the discrete-time systems as rigorously shown in this paper.  
Moreover, we prove a manifold (subspace) turnpike property of the discrete-time optimal control problem that provides the foundation of (economic) predictive controllers~\cite{FaulMull18}.

This work is organized as follows: In Section \ref{se:prelims}, we recall existing results about strict dissipativity of the Continuous-time Optimal Control Problem (CT-OCP) and define the notion for the corresponding Discrete-time Optimal Control Problem (DT-OCP). In Section~\ref{se:implicit}, we elaborate issues with using an implicit mid-point discretization method to preserve strict dissipativity. 
In Section~\ref{se:DDR}, we recall the formulation of DDR for discrete-time pH systems. We use DDR to establish strict dissipativity and turnpike properties for DT-OCP for nonlinear pH systems with quadratic Hamiltonians in Section~\ref{se:dissipativity_ddr}. Section \ref{se:num_examp} contains two numerical examples that validate our findings, before conclusions are drawn in Section~\ref{se:conclusion}.\medskip

\noindent {\bf Notation:} 
For a matrix $X\in \mathbb{R}^{n\times n}$, $X > 0$ $(\geq 0)$ denotes positive (semi)-definiteness.
The space of measurable and absolutely integrable functions is denoted by  $L^1(0,T;\rline)$ and $W^{1,1}(0,T;\rline^n)$ denotes the space of functions in $L^1(0,T;\rline^n)$ with weak derivative in $L^1(0,T;\rline^n)$. The interior of a set \(\mathcal{S}\subset \mathbb{R}^n\) is defined as \(\mathrm{int}(\mathcal S)\). For a manifold \(\mscr \subseteq \rline^n\), we set $\mathrm{unpp}(\mscr) := \{x \in \rline^n \mid\ \exists\; \text{unique}\; \xi \in \mscr\; ,\; \dist(x,\mscr) = \|x - \xi\|\}$ and we abbreviate $[0:N]=\{0,1,\ldots,N\}$. For a map \(f : \mathbb{R}^n \to \rline\), \(Df_x\) defines the differential of \(f\) at \(x \in \rline^n\).

\section{Preliminaries and problem description} \label{se:prelims}
In this part, we introduce the motivation and the problem of this work. To this end, let us consider a {linear} pH system %
\begin{align}\label{eq:phODE}
	\dot x(t) = (J-R)Qx(t) + Bu(t), \quad y(t) = B^\top Q x(t),
\end{align}
where $J=-J^\top\in\rline^{n\times n}$ is a skew-symmetric interconnection matrix, $0 \leq R=R^\top \in \rline^{n\times n}$ models the dissipation, $0 < Q=Q^\top\in\rline^{n\times n}$ corresponds to the Hamiltonian energy and $B\in\rline^{n\times m}$. For time horizon $T>0$, straightforward computations yield %
the energy balance
\begin{align}\label{eq:energy_bal}
	\!H(x(T)) \!-\! H(x(0)) =\! \int\nolimits_0^T \!\!\!\!u(t)^\top \!y(t) - \|R^{\nicefrac12}Qx(t)\|_2^2 \,\mathrm{d}t
\end{align}
with Hamiltonian $H(x):=\frac12 x^\top Q x$ and Cholesky factor $R^{\nicefrac12}$ of $R$. Whereas the left-hand side resembles the energy change, the right-hand side consists of the dissipated energy and the supplied energy given by the product of input and output; stating a natural quantity to minimize. In view of this energy balance, minimal-energy control problems in continuous time have been formulated and extensively studied in recent years, see~\cite{Schaller21,Karsai2024,Faulwasser2022a,AronMehr24}. For other tasks achieved by optimizing pHs, we refer to the learning-based approaches \cite{koelsch2021optimal,massaroli2022optimal} and to \cite{stegink2017port,gernandt2025port} for optimization-based control by interconnection. 

In applications, a control task could be an optimal set point transition:
Steer a system from $x_0\in \mathbb{R}^n$ to $x_T\in \mathbb{R}^n$ in time $T>0$ with minimal energy supply taking into account the control constraints $u(t)\in \mathbb{U}, t\in [0,T]$. In continuous time, this task can be addressed by solving the following optimal control problem (OCP):
\begin{align}\label{eq:ct_ocp}\tag{CT-OCP}
	\begin{split}
		&\min_{u \in L^1(0,T;\uscr)} \int\nolimits_0^T u(t)^\top y(t)\,\mathrm{d}t \\
		&\mathrm{s.t.}\, \ \dot x(t) = (J-R)Qx(t) + Bu(t),\\
		&\phantom{\mathrm{s.t.}}\, \ x(0) = x_0, \ x(T) = x_T, y(t)=B^\top Qx(t) 
	\end{split}
\end{align}
We denote the optimal control by $u^*\in L^1(0,T;\uscr)$ and the associated optimal states by $x^*\in W^{1,1}(0,T;\rline^n)$.

In view of inverse optimality we note that up to now it is an open problem whether the optimal control problem \eqref{eq:ct_ocp} admits a feedback-type solution, i.e., it is in particular unclear if any passivity-based control schemes such as damping injection are indeed energy optimal in the above sense.
\begin{proposition} \label{pr:Schaller_21}
	OCP~\eqref{eq:ct_ocp} is strictly dissipative w.r.t. the subspace $\mathcal{M} = \operatorname{ker}R^{\nicefrac12} Q$, if there exists a storage function $S:\rline^n\to [0,\infty)$ and $\alpha\in\kscr_\infty$\footnote{A continuous function $\alpha:\mathbb{R}_{\geq 0} \to \mathbb{R}_{\geq 0}$ is said to be of class $\mathcal{K}_\infty$ if $\alpha(0)=0$ and if $\alpha$ is strictly increasing and unbounded.} such that
	\begin{align*}
		S(x(T)) \!-\! S(x(0)) \!\leq\! \int_0^T \!\!\! u^*(t)^\top y^*(t) \!-\!\alpha(\dist(x^*(t), \mathcal{M}))\,\mathrm{d}t
	\end{align*}
	holds for all optimal pairs $u^*$ and $x^*$.
\end{proposition}
Strict dissipativity is a central building block in the formulation and analysis of MPC schemes in~\cite{ZanoFaul18, book:Grune2017}. 
In particular, the close relation to the turnpike property~\cite{GrunMull16} ensures that optimal trajectories on long horizons approach (in the simplest case) an optimal steady state. 
In~\cite{Schaller21}, Proposition \ref{pr:Schaller_21} is used to establish the turnpike behavior of the OCP w.r.t.\ the conservative subspace $\operatorname{ker}R^{\nicefrac12}Q$ for continuous-time pH systems. For manifold dissipativity in the context of trims of mechanical systems, we refer to \cite{FaulFlass22} and in the context of energy-optimal control of pH descriptor systems, see \cite{Karsai2024}.
To enable predictive control using time discretizations, it is crucial to obtain this dissipativity and turnpike behavior also for the discrete-time case.

We now define strict dissipativity w.r.t\ a manifold for \emph{nonlinear} and \emph{discrete-time} optimal control problems. 
To this end, let $\xscr\subset \rline^n$ and $\uscr\subset \rline^m$ be the set of admissible state and input values. For $N\geq 1$ and initial state $x(0)=x_0\in \xscr$, $\uscr^N(x_0)= \{(u_k)_{k=0}^{N-1} \subset \uscr \mid x(k,x_0)\in\xscr, k \in [1:N] \}$ is defined, where $x(k, x_0)$ is the state at time $k \in [0:N]$ subject to $x(0,x_0) = x_0$ and application of $u=(u_k)_{k=0}^{N-1}$. 

\begin{definition}\label{de:strict_diss_disc} %
	For $x_0,x_N\in \mathbb{X}$, the discrete-time OCP 
	\begin{align}\label{eq:dt_ocp}\tag{DT-OCP}
		\begin{split}
			&\min_{u\in\uscr^N(x_0)}\sum\nolimits_{k=0}^{N-1} l(x_k,u_k) \\
			&\mathrm{s.t.}\, \ x_{k+1} = f(x_k,u_k),\ y_k = g(x_k,u_k), \\ 
			&\phantom{\mathrm{s.t.}}\, \ x(0,x_0) = x_0, \ x(N,x_0) = x_N.
		\end{split}
	\end{align}
	is said to be strictly dissipative w.r.t.\ a manifold $\mathcal{M}\subset\rline^n$ if there exists storage function $S : \xline \to [0,\infty)$ and $\mathcal{K}_\infty$-function~$\alpha$ such that all optimal control sequences $u^*\coloneqq (u_k^*)_{k=0}^{N-1}$ and associated optimal state $x^* \coloneqq (x^*_k)_{k=0}^{N}$ satisfy, for each $k\in [0:N-1]$, the dissipation inequality 
	\begin{equation}\label{eq:str_disp_subsp_K}
		S(x^*_{k+1})-S(x^*_k) \leq l(x_k^*,u_k^*) - \alpha(\dist(x^*_k,\mathcal{M})).
	\end{equation}
\end{definition}
We note that a summation of \eqref{eq:str_disp_subsp_K} implies
\begin{equation}\label{eq:str_disp_subsp_DT}
	S(x_N)-S(x_0) \leq \sum\nolimits_{k=0}^{N-1} l(x_k^*,u_k^*) - \alpha(\dist(x^*_k,\mathcal{M})).
\end{equation}
\noindent %

While energy-preserving discretization schemes are well-established in the pH literature~\cite{kotyczka2019, MehrMora19}, discretizations or discrete-time formulations that explicitly preserve strict dissipativity properties of OCPs remain unexplored up to now. Hence, in this work we address the following problem:
\begin{problem}
Find an appropriate discrete-time representation (DT-OCP) of the optimal control problem (CT-OCP) that preserves strict dissipativity in the sense of Definition~\ref{de:strict_diss_disc}.
\end{problem}

\section{Implicit-midpoint discretization} \label{se:implicit}

\noindent In this section, we show that preserving strict dissipativity is non-trivial and may not be straightforwardly achieved by using standard energy-preserving discretization schemes. %
To keep the discussion simple, in Section~\ref{se:implicit}, we focus on linear pH systems and the implicit midpoint scheme, i.e., one of the most used and simplest dissipativity-preserving schemes for pH systems \cite{kotyczka2019,MehU22}.
To this end, we introduce the implicit midpoint discretization of \eqref{eq:phODE}\footnote{For simplicity, we choose a step size $h = 1$, as we are particularly interested in strict dissipativity for all and not only \textit{small} time steps.}
\begin{align}\label{eq:midpoint}
[ I-\tfrac{1}{2}(J-R)Q ] x_{k+1} = [ I+\tfrac{1}{2}(J-R)Q ] x_k + Bu_k.
\end{align}
Before providing the main result, we briefly provide a result showing invertibility due to the pH structure.
\begin{lemma}\label{le:J_invertibility}
Let $Q,R\in \rline^{n\times n}$ be symmetric positive semidefinite and $J\in \rline^{n\times n}$ be skew-symmetric. Then, the matrix $J_- := I-\tfrac{1}{2}(J-R)Q$ is invertible.
\end{lemma}
\begin{proof}
Denote $V = \ker Q$ and let $\rline^n = V \oplus V^\perp$. For any non-zero $z\in V$, $J_-z = z\neq 0$. For any $z\in V^\perp$, we have $\langle z,QJ_- z\rangle = \|z\|^2_Q + \tfrac12\|R^{\nicefrac{1}{2}}Qz\|^2>0$, which implies that $J_-z \not\in V$. Hence, $J_-z\neq 0$ for all non-zero $z\in \rline^n$ and invertibility of $J_-$ follows.    
\end{proof}

Denote $\Delta H_{k,k+1} \coloneqq H(x_{k+1})-H(x_k)$. The energy balance for \eqref{eq:midpoint} is $\Delta H_{k,k+1} =  \tfrac{1}{2}(x_{k+1}^\top Qx_{k+1}-x_k^\top Q x_k)$ which, upon using Lemma \ref{le:J_invertibility}, can be rewritten as
\begin{eqnarray}
\Delta H_{k,k+1} & = & \tfrac{1}{2}(x_{k+1}-x_k)^\top Q (x_{k+1}+x_k) \nonumber \\    
& = & u^\top_k y_k - \tfrac{1}{4} \| R^{\frac{1}{2}}QJ_-^{-1}[ 2x_k+Bu_k ]\|^2_2 \label{eq:energy_balance_linear_DTpH}
\end{eqnarray}
with $y_k = \frac{1}{2}B^\top QJ_-^{-1}(2x_k+Bu_k)$ for all $k\in [0:N-1]$, where $J_-$ is defined in Lemma \ref{le:J_invertibility}, $J_+ := I+\tfrac{1}{2}(J-R)Q$ and we use $x_{k+1}+x_k = J_-^{-1}(2x_k+Bu_k)$ and $x_{k+1}-x_k = \tfrac{1}{2}(J-R)QJ_-^{-1}(2x_k+Bu_k) + Bu_k$. Note that this output differs from the output of \eqref{eq:phODE}. However, our choice above is consistent with the output obtained by discrete gradient method (as defined in Section \ref{se:DDR}), and hence crucial to obtain the energy balance. Using the implicit midpoint scheme, a natural discrete-time analogue of \eqref{eq:ct_ocp} is obtained with $l(x_k,u_k) = u_k^\top y_k$ in \eqref{eq:dt_ocp}
\begin{align}
& \min_{u_k\in\uscr_N(x_0)}\, \ \sum\nolimits_{k=0}^{N-1}u^\top_k y_k \nonumber \\
& \mathrm{s.t.} \, \ x_{k+1} = J_-^{-1}J_+x_k + J_-^{-1}Bu_k, \label{eq:linearDTpHOCP} \\
& \phantom{\mathrm{s.t.}} \, \ x(0) = x_0, x(N) = x_N, y_k = \frac{1}{2}B^\top QJ_{-}^{-1}(2x_k+Bu_k). \nonumber
\end{align}

Next, under a particular assumption on the optimal controls, we show the strict dissipativity property w.r.t.\ a subspace for the DT-OCP in~\eqref{eq:linearDTpHOCP}. 
\begin{proposition}\label{pr:mid_pt_result}
The optimal control problem in \eqref{eq:linearDTpHOCP} is strictly dissipative w.r.t the subspace  $\operatorname{ker} R^{\nicefrac{1}{2}}QJ_-^{-1}$ with storage function $S=H$ if and only if $Bu^*_k\in \ker R^{\nicefrac{1}{2}}QJ^{-1}_{-}$ for all optimal controls $(u^*_k)_{k=0}^{N-1}$.
\end{proposition}
\begin{proof}
Suppose that $Bu^*_k\in \ker R^{\nicefrac{1}{2}}QJ^{-1}_{-}$ for all optimal control sequences $u_k^*$. Then, \eqref{eq:energy_balance_linear_DTpH} implies
\begin{align*}
	H(x_N)-H(x_0) = \sum\nolimits_{k=0}^{N-1} \left(u_k^{*^{\top}} y^*_k- \tfrac{1}{4} \| R^{\nicefrac{1}{2}}QJ_-^{-1}x^*_k \|_2^2  \right).
\end{align*}
Using \cite[Lemma 13]{Schaller21} with $R^{\nicefrac{1}{2}}QJ_-^{-1}$ in place of $R^{\nicefrac{1}{2}}Q$, \eqref{eq:linearDTpHOCP} implies strict dissipativity w.r.t. subspace $\operatorname{ker}R^{\nicefrac{1}{2}}QJ_-^{-1}$. Next, suppose that the strict dissipativity condition holds for \eqref{eq:linearDTpHOCP} and, seeking a contradication, assume there is an optimal control $(u_k^*)_{k=0}^{N-1}$ such that $Bu_k^*\not\in \operatorname{ker}R^{\nicefrac{1}{2}}QJ_-^{-1}$ for some $k\in[0:N-1]$. In \eqref{eq:linearDTpHOCP}, fix $N = 1$, let $x_0\in\operatorname{ran} B$ and $x_0\not\in\operatorname{ker}R^{\nicefrac{1}{2}}QJ_-^{-1}$. Note that such an $x_0$ exists, unless $\operatorname{ran} B\subset \operatorname{ker}R^{\nicefrac{1}{2}}QJ_-^{-1}$ in which case the condition on $u_k^*$ is already satisfied. Fix $x_N = -x_0$. Under these conditions, the control sequence that solves \eqref{eq:linearDTpHOCP} is given by $Bu_0^* = -2x_0$ (which is the only admissible control), which is also optimal with cost zero as $y_0=0$. Using  \eqref{eq:str_disp_subsp_K}, with $S = H$, $l(x_k^*,u_k^*)=u_k^\top y_k$, $\mathcal{V}=\operatorname{ker}R^{\nicefrac{1}{2}}QJ_-^{-1}$, and \eqref{eq:energy_balance_linear_DTpH} for $k=0$
$$\alpha(\dist(x_0,\operatorname{ker}R^{\nicefrac{1}{2}}QJ_-^{-1})) \leq \tfrac{1}{4}\|R^{\nicefrac{1}{2}}QJ_-^{-1}(2x_0+Bu^*_0)\|^2_2$$
must hold, which is a contradiction since $2x_0+Bu_0^*=0$ and $x_0\not\in\operatorname{ker}R^{\nicefrac{1}{2}}QJ_-^{-1}$.  
\end{proof}
From Proposition \ref{pr:mid_pt_result}, we observe preservation of the strict dissipativity property of (CT-OCP) in the discretized version of the OCP if $Bu^*_k\in \operatorname{ker} R^{\frac{1}{2}}QJ^{-1}_{-}$ holds. However, this condition is typically hard to check %
as it depends on the optimal control. Moreover, when $R>0$ and $\operatorname{ker} R^{\nicefrac{1}{2}}QJ_-^{-1} = \{0\}$, the applicability of Proposition \ref{pr:mid_pt_result} is limited as it does not capture the strict dissipativity behavior under non-zero optimal inputs.
In the following, we show how dissipativity may be ensured without such limiting conditions by using DDRs. Several other discretization schemes, see \cite{XiAnPa:17}, \cite{MoBiAs:25} and the references therein, could also be investigated in future research.
\section{Difference-differential Representation} \label{se:DDR}
\noindent In this section, we recall the difference and differential representation (DDR; \cite{Moreschini2021Thesis}) for discrete-time nonlinear pH systems. To circumvent a well-posedness issue in the standard passivity notion, the DDR utilizes the idea of $u$-average outputs along with discrete gradient functions to define $u$-average passivity~\cite{Monaco1998}.
We first  state the notion of discrete gradient, see \cite[Definition 2.1.1]{Moreschini2021Thesis}: The discrete gradient of a differentiable function $H:\rline^n\to\rline_{\geq 0}$ is $\bar\nabla H:\rline^n\times \rline^n\to \rline^n $ defined as follows: For $v , w \in \rline^n$, the $i$-th entry of $\bar\nabla H(v,w)\coloneqq\left.\bar{\nabla}H\right|_{v}^{w} $ is defined by by the expression
\begin{equation*}\label{eq:discrete_grad_def}
\frac{1}{(w_i-v_i)} \int_{v_i}^{w_i} \frac{\partial H(v_1,\ldots v_{i-1},\xi, w_{i+1},\ldots , w_n)}{\partial \xi} d\xi 
\end{equation*}
such that $(w-v)^\top \bar{\nabla } H(v,w) = H(w)-H(v)$ and $\bar{\nabla} H(v,v) = \nabla H(v)$
hold. A constructive form of discrete gradient can be provided \cite[Lemma 2.1.1]{Moreschini2021Thesis} by the mean value theorem as $\bar{\nabla}H|_{v}^{w} = \int_0^1 \left.\nabla H\right|_{v+s(w-v)} ds$.

Using the notion of discrete gradient function, we now state DDR of pH system with constant input map $B$, see also \cite[Definition 4.1.1]{Moreschini2021Thesis} for more details. From now on, we focus here on the single input case, i.e., $m=1$ and consider the nonlinear counterpart of \eqref{eq:ct_ocp} replacing $J$ and $R$ by state-dependent matrices and $Q$ by an energy gradient $\nabla H$.
\begin{definition}\label{de:DDR_def} %
Let $J:\rline^n\to \rline^{n\times n}$ be pointwise skew-symmetric, $R: \rline^n\to \rline^{n\times n}$ pointwise symmetric and positive semidefinite and $H: \rline^n \to \rline_{\geq 0}$ be the Hamiltonian function. 
Then, the DDR of a discrete-time pH system over $\rline^n$ in map form for input $u_k\in\rline$ is given by
\begin{subequations}\label{eq:ddr_Z}
	\begin{align}
		x_{k+1}^\aut &= x_k + (J(x_k)-R(x_k))\left.\bar{\nabla}H\right|_{x_k}^{x_{k+1}^{\aut}}, \label{eq:implicit_ddr_Z}  \\ 
		x_{k+1} &= x_k + (J(x_k)-R(x_k))\left.\bar{\nabla}H\right|_{x_k}^{{ x_{k+1}^{\aut}}} + Bu_k, \label{eq:DT-pH map form} \\
		Y_k &= B^\top \left.\bar{\nabla}H\right|_{{ x_{k+1}^{\aut}}}^{x_{k+1}} \label{eq:DT-pH_map_output},
	\end{align}
\end{subequations}
where, at time instant $k\in\nline$, $x_{k+1}^{\aut}$ is the autonomous  (unforced) one-step ahead evolution,  i.e., equal to $x_{k+1}$ if $u_k =0$, $x_{k+1}$ is the controlled one-step ahead evolution. 
\end{definition} 
The DDR of the pH system~\eqref{eq:ddr_Z} comprises 
(i) an implicit autonomous pH dynamics that generates the intermediate one-step ahead solution~$ x_{k+1}^{\aut}$ from~$x_k$ and 
(ii) a differential equation that models the effect of $u$ and yields the state $x_{k+1}$ using state $ x_{k+1}^{\aut}$  as the initial condition and input~$u_k$. Following the discussion around \cite[Theorem 4.1.1]{Moreschini2021Thesis}, it can be easily shown that the DDR of the pH system satisfies the energy balance equation
\begin{small}
\begin{align}\label{eq:DT-pH_energybal}
	\underbrace{\Delta H_{k,k+1}}_{\text{stored energy}} = \underbrace{u_k B^\top \!\!\left.\overline{\nabla}H\right|_{{ x_{k+1}^{\aut}}}^{x_{k+1}}}_{\text{ supplied energy}} \!-\! \underbrace{\left.\overline{\nabla}H^\top\right|_{x_k}^{{x_{k+1}^{\aut}}}\!\!\!R(x_k)\left.\overline{\nabla}H\right|_{x_k}^{{ x_{k+1}^{\aut}}}}_{\text{dissipated energy}}. 
\end{align}
\end{small}
Note that the total dissipated energy depends on the dissipation matrix $R$ and is independent of the input $u$  while the total supplied energy depends on $u$ and the discrete gradient computed between $x_{k+1}^{\aut}$ and $x_{k+1}$, similar to the energy balance for continuous-time dissipative systems. This makes DDR a suitable candidate to analyze dissipativity and turnpike properties, as will be evident in Section \ref{se:dissipativity_ddr}. 

Although \eqref{eq:ddr_Z} comprises an implicit equation in terms of $x_{k+1}^{\aut}$, in \cite[Proposition 4.1.2]{Moreschini2021Thesis} it has been established that for a class of pH systems with quadratic Hamiltonian, it is possible to obtain an explicit form of \eqref{eq:ddr_Z} under invertibility assumption on $J_-(x) := I-\tfrac{1}{2}(J(x)-R(x))Q$. Below, we first prove the invertibility of  $J_-(x)$ and then summarize the aforementioned result in a proposition.
\begin{lemma}\label{le:J_inverse_nonlinear}
For $x\in\rline^n$, suppose that  $J:\rline^n\to \rline^{n\times n}$ is pointwise skew-symmetric and $R: \rline^n\to \rline^{n\times n}$ is pointwise symmetric, and $0 \leq Q=Q^\top \in \rline ^n$.  Then, for all $x\in \rline^{n}$ the matrix $J_-(x)$ is invertible.   
\end{lemma}
\begin{proof}
The proof is same as the proof of Lemma \ref{le:J_invertibility} applied pointwise for all \(x \in \rline^n\).
\end{proof}

The following proposition provides a DDR of nonlinear discrete-time port Hamiltonian systems with quadratic Hamiltonian in explicit form, see also \cite[Proposition 4.1.2]{Moreschini2021Thesis}. 
In particular, Lemma \ref{le:J_inverse_nonlinear} enables us to remove the invertibility assumption w.r.t.\ \(J_-(x)\) from \cite[Proposition 4.1.2]{Moreschini2021Thesis}.
\begin{proposition}\label{pr:DDR_quadratic_Ham}
Given a pH system in DDR \eqref{eq:ddr_Z} with Hamiltonian $H(x) =\tfrac{1}{2}x^\top Qx$, $0 < Q=Q^\top \in \rline^{n\times n}$ and the discrete gradient function $\bar\nabla H|_w^v = \tfrac{1}{2}Q(w+v)$, 
the associated autonomous dynamics in explicit form is $x^{aut}_{k+1} = J_-^{-1}(x_k)J_+(x_k)x_k$ and the controlled dynamics is
\begin{equation} \label{eq:DDR_quadratic_ham_controlled}
	x_{k+1} =  J_-^{-1}(x_k)J_+(x_k)x_k + Bu_k 
\end{equation}
with $u$-average passive output given by \eqref{eq:DT-pH_map_output}, where 
$J_+(x) := I+\frac{1}{2}(J(x)-R(x))Q$.
\end{proposition}
The explicit DDR representation given by Formula~\eqref{eq:DDR_quadratic_ham_controlled} is easily implementable and renders the following approach computationally tractable. 
\section{Dissipativity and turnpike via DDR} \label{se:dissipativity_ddr}
\noindent In this section, we use the differential-difference representation (DDR) of Proposition \ref{pr:DDR_quadratic_Ham} to deduce a discrete-time energy-optimal control problem that is strictly dissipative. %
As a consequence, we establish the turnpike behavior of the optimal state trajectories for the discretized optimal control problem. %
Below, we define the DT-OCP for a nonlinear pH system with a quadratic Hamiltonian $H(x) = \frac{1}{2}x^\top Q x$. 
\begin{align}
\min_{u_k} 
&\sum\nolimits_{k=0}^{N-1}l(x_k,u_k) \ \mathrm{s.t.} \ \text{Eq.}~\eqref{eq:DDR_quadratic_ham_controlled} \ \text{holds}\   \nonumber\\ 
\mathrm{with} \ \ & x(0) = x_0, \, x(N) = x_N, \label{eq:linearDTpHOCP_average} \\
\ \ & Y_k(x_k,u_k) = B^\top QJ_-^{-1}{ (x_k)}J_+{ (x_k)} x_k + \frac{1}{2}B^\top QBu_k. \nonumber
\end{align}
Let $l(x_k,u_k)=u_kY_k$, then the energy balance in \eqref{eq:DT-pH_energybal} implies
\begin{small}
\begin{align}\label{eq:energy-balance-sum}
	\begin{split}
		\sum_{k=0}^{N-1}\!l(x_k,u_k) &= \Delta H_{0,N} \!+\! \sum_{k=0}^{N-1}\Big\|R(x_k)^{\nicefrac{1}{2}}QJ_{-}^{-1}(x_k)x_k\Big\|_2^2.
	\end{split}
\end{align}
\end{small}
\noindent We now state the main assumption of this section, see \cite{Karsai2024}.
\begin{assumption}\label{asm:manifold}
Let the following conditions hold for the set $\mscr := \{x_k \in \rline^n \mid R(x_k)^{\nicefrac{1}{2}}QJ^{-1}_-(x_k)x_k = 0\}$:\\
1) The map \(x_k \mapsto R(x_k)^\frac{1}{2}QJ^{-1}_-(x_k)x_k\) is of class \(C^2\).\\
2) The set \(\mscr\) is nonempty and there exists an open neighborhood \(G \subseteq \rline^n\) of \(\mscr\) and $0<s<n$ such that for all \(x_k \in G\), it holds that \(\dim (\ker D(R(x)^\frac{1}{2}QJ(x)^{-1}_-x)_{\vert x = x_k}) = s\).\\
3) Let \(V \subseteq \rline^n\), \(\mscr \subseteq V \subseteq \mathrm{int} (\mathrm{unpp}(\mscr))\), be an open set, where the estimate of \cite[Lemma 5]{Karsai2024} holds. Any optimal trajectory \(x^\ast_k\) of the DT-OCP \eqref{eq:linearDTpHOCP_average} remains in \(V\) for all times.
\end{assumption}
We now show the strict dissipativity of the minimal energy problem in \eqref{eq:linearDTpHOCP_average} with respect to a manifold.

\begin{theorem}\label{th:DTOCP_strict_dissip}
Consider Assumption \ref{asm:manifold} is satisfied. Then, the discrete-time optimal control problem in \eqref{eq:linearDTpHOCP_average} is strictly dissipative with respect to the manifold $\mscr := \{x_k \in \rline^n \mid R(x_k)^{\frac{1}{2}}QJ^{-1}_-(x_k)x_k = 0\}$ with storage function \(H\). 
\end{theorem}
\begin{proof}
As Assumption \ref{asm:manifold} is satisfied, invoking \cite[Lemma 5]{Karsai2024}, there exists an open set \(V \subseteq \rline^n\) and a scalar constant \(c > 0\) such that \(\mscr \subseteq V \subseteq \mathrm{int}(\mathrm{unpp(\mscr)})\) and 
\begin{align}\label{eq:estimate_manifold}
	c \; \dist(x_k, \mscr) \leq \|R(x_k)^{\frac{1}{2}}QJ_-^{-1}(x_k)x_k\| \quad\forall\,x_k \in V. 
\end{align}
So, \eqref{eq:estimate_manifold} is true along any optimal trajectory \(x^\ast_k\) of the DTOCP \eqref{eq:linearDTpHOCP_average}. For an optimal control \(u^\ast_k\), associated trajectory \(x^\ast_k\) and output \(Y_{k}^\ast(x^\ast_k,u^\ast_k)\), \eqref{eq:energy-balance-sum} is satisfied. 
Then, setting $\alpha : s \mapsto c^2 s^2 \in \mathcal{K}_\infty$,  using \eqref{eq:estimate_manifold}, yields the claim via $\Delta H_{0,N} \leq \sum_{k=0}^{N-1} u^\ast_k Y^\ast_{k}(x^\ast_k,u^\ast_k) - \alpha(\dist(x^\ast_k,\mscr)). $
\end{proof}
\begin{remark}\label{re:sampling}
For simplicity of presentation, in this work, we fix an uniform step size $h = 1$ to obtain the DT-OCP in \eqref{eq:linearDTpHOCP} and \eqref{eq:linearDTpHOCP_average}.  However, our results hold true for a general step size $h\in(0,1)$ as well. For sufficiently small step-size $h\in (0,1)$, leveraging convergence of the numerical scheme and continuity, it is clear that the strict-dissipativity is preserved with both implicit scheme and DDR. However, for larger step sizes, while DDR still preserves strict dissipativity, the implicit scheme fails as will be illustrated by our numerical examples in Section~\ref{se:num_examp}. Furthermore, the result presented here also apply to general discrete-time pH systems, including the ones obtained via exact discretization (equivalent sampled-model) \cite{MoMoNo:19}, \cite{MoNoTi:11}.
\end{remark}

Before moving further, consider the steady-state optimal control problem corresponding to \eqref{eq:linearDTpHOCP_average}
\begin{align}\label{eq:DTOCP_steady_state}
& \min_{\bar{u} \in \mathbb{U}} l(\bar x,\bar u):= \bar{u}^\top B^\top Q J_-^{-1}(\bar{x})\bar{x} + \tfrac{1}{2}\bar{u}^\top B^\top Q B \bar{u} \nonumber\\
\mathrm{s.t.}\; 0 &= (J(\bar{x})-R(\bar{x}))QJ^{-1}_-(\bar{x})\bar{x} + B\bar{u}.
\end{align}
The following proposition plays an important role to prove the turnpike result afterwards.
\begin{proposition}
Consider a steady state DT-OCP \eqref{eq:DTOCP_steady_state}. Then 
\begin{align}\label{eq:cost_function_steady state}
	l(\bar x,\bar u) = \|R^\frac{1}{2}(\bar x)QJ_-^{-1}(\bar x)\bar x\|^2 \geq 0,
\end{align}
holds for the optimal value. Moreover, it is zero if and only if 
$\mathcal{S} := \{(\bar x, \bar u) \in \mscr \times \mathbb{U} \mid B\bar u = -J(\bar x)QJ_-^{-1}(\bar x)\bar x\} \neq \emptyset$.
\end{proposition}
\begin{proof}
At steady state, we have \(x_k = x_{k+1} = \bar x\) which implies  \(H(x_k)=H(x_{k+1})=H(\bar x)\). Then, $l(\bar x,\bar u) = \|R^{\nicefrac{1}{2}}(\bar{x}_k)QJ_-^{-1}(\bar{x}_k)\bar{x}_k\|^2 \geq 0$. %
For the second claim, assume that there exists a steady state \((\bar x, \bar u) \in \mathcal{S}\). So, from \eqref{eq:DTOCP_steady_state}, we have that \(R(\bar x)QJ_-^{-1}(\bar x)\bar x = 0\),  which implies \(l(\bar x, \bar u) = 0\). Conversely, let \((\bar x, \bar u)\) be such that \(l(\bar x, \bar u) = 0\). From \eqref{eq:cost_function_steady state}, we get \(R^{\frac{1}{2}}(\bar x)QJ_-^{-1}(\bar x)\bar x = 0\) and so \(\bar x \in \mscr\). From \eqref{eq:DTOCP_steady_state} we get \(B\bar u = -J(\bar x)QJ_-^{-1}(\bar x)\bar x\), i.e., \(\mathcal{S}\) is non-empty. 
\end{proof}
\begin{definition}
We say that a general discrete-time optimal control problem \eqref{eq:dt_ocp} has the \textit{sum turnpike property} on a set \(\mathcal{I}_{tp} \subset \xline\) with respect to a manifold \(\mscr \subset \rline^n\) if for all compact \(X_0 \subset \mathcal{I}_{tp}\) there is a constant \(C > 0\) such that for all \(x_0 \in X_0\), and all \(N > 0\), each optimal pair \((x^\ast, u^\ast)\) of the OCP with \(x^\ast_0 = x_0\) satisfies
\begin{align}\label{eq:sum-tunpike-property}
	\sum\nolimits_{k=0}^{N-1} \dist^2(x^\ast_k, \mscr) \leq C {.}
\end{align}
\end{definition}
\begin{remark}
The interpretation of the estimate \eqref{eq:sum-tunpike-property} is as follows: As the upper bound $C\geq 0$ is uniform in the horizon $N$, increasing the horizon implies that the majority of the summands is small, thus implying that $x^*_k$ is close to $\mathcal{M}$ for the majority of the time instances. This also implies measure turnpike property with respect to the manifold $\mscr$.
\end{remark} 
\begin{assumption}\label{asp:compact_trajectory}
For any compact set of initial conditions \(X_0 \subset \xline\), there is a compact set \(K \subset V \subset \xline\) such that for all horizons \(N\), any optimal state trajectory of \eqref{eq:linearDTpHOCP_average} for all \(x_0 \in X_0\) and for horizon \(N\) is contained in \(K\).
\end{assumption}
\noindent Before proving a turnpike result, we define the set 
\begin{align*}
& \cscr(\mscr_{opt}, x_N) := \{x_0 \in \rline_n \mid \exists\, k_1, k_2 \in \mathbb{N}, (u^1_k)_{k=1}^{k_1}, (u^2_k)_{k=0}^{k_2}, \\
& \hspace*{.8cm} \bar{x} \in \mscr_{opt}: x(k_1,u_1,x_0) = \bar x, x(k_2+1,u_2,\bar x) = x_N \}
\end{align*}
with \(\mscr_{opt} := \{\bar x \in \mscr \mid \exists\; u \in \mathbb{U}\; \text{with}\; (\bar x, \bar u) \in \sscr\}\).
\begin{theorem}\label{thm:sum_turnpike}
Consider Assumption \ref{asm:manifold}  and \ref{asp:compact_trajectory} hold and \(\cscr(\mscr_{opt},x_N) \neq \emptyset\). Then the DTOCP \eqref{eq:linearDTpHOCP_average} has sum turnpike property on the set \(\mathcal{I}_{tp} = \cscr(\mscr_{opt},x_N)\) with respect to \(\mscr\).
\end{theorem}
\begin{proof}
Let \(X_0 \subset \mathcal{I}_{tp}\) be compact, \(x_0 \in X_0\), and \((x^\ast, u^\ast)\) be an optimal pair for \eqref{eq:linearDTpHOCP_average}. For any $(u_k)_{k=0}^{N-1}$%
and corresponding trajectory \(x(\cdot, u, x_0)\), we have \(\sum_{k=0}^{N-1}l(x^\ast,u^\ast) \leq \sum_{k=0}^{N-1}l(x_k,u_k)\) due to optimality.

Now, for \(N \leq k_1+k_2-1\), the sum turnpike property \eqref{eq:sum-tunpike-property} is trivially satisfied with \(C := (k_1+k_2) \sup \{\dist (x,\mscr) \mid x \in K\}\). For, \(N \geq k_1+k_2\), define
\begin{small}
	\begin{equation}\label{eq:control_construction}
		u_k := \begin{cases}
			u^1_k & k\in [0 : k_1-1],\\
			\bar u & k \in [k_1 : N-k_2-1],\\
			u^2_{k - (N-k_2)} & k \in [N-k_2 : N-1],
		\end{cases}
	\end{equation}
\end{small}
where \(\bar u\) is the steady state control such that \((\bar x, \bar u) \in \sscr\), i.e., the corresponding stage cost is zero. Hence, we have $|\sum_{k=0}^{N-1}l(x_k,u_k)| = |\sum_{i=1}^{2}\sum_{k=0}^{k_i-1} l(x_k^i,u_k^i)|\leq C_1$ for $C_1\geq 0$ independent of $N$, where $(x_k^i)_{k\in [0: t_i-1]}$ is the trajectory corresponding to $(u_k^i)_{k\in k\in [0:k_i-1]}$, $i=1,2$. Then, from \eqref{eq:energy-balance-sum} and \eqref{eq:estimate_manifold}, we have 
\begin{align}
	\sum_{k=0}^{N-1} \dist^2(x^\ast,\mscr) \leq \underbrace{\frac{C_1}{c^2} - \frac{1}{c^2}[H(x^\ast(N)) - H(x_0)]}_{:=C},
\end{align}
where $c > 0$ is a constant as defined in \eqref{eq:estimate_manifold}. As \(H\) is continuous and Assumption \ref{asp:compact_trajectory} is satisfied, the right-hand side can be estimated, which is independent of horizon \(N\). %
\end{proof}
\section{Numerical examples} \label{se:num_examp}
\noindent In this section, we present two numerical examples to confirm our theoretical findings. To this end, we solve \eqref{eq:dt_ocp} for two different problems utilized in previous works, considering the dissipativity. 

\begin{figure}[htb]
\centering   
\includegraphics[width=.45\textwidth]{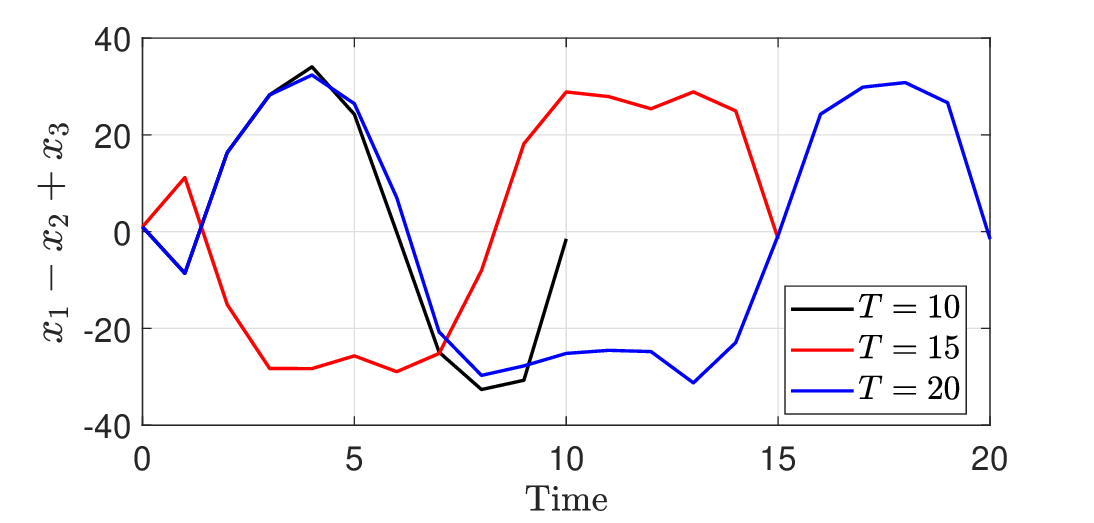} 
\includegraphics[width=.45\textwidth] {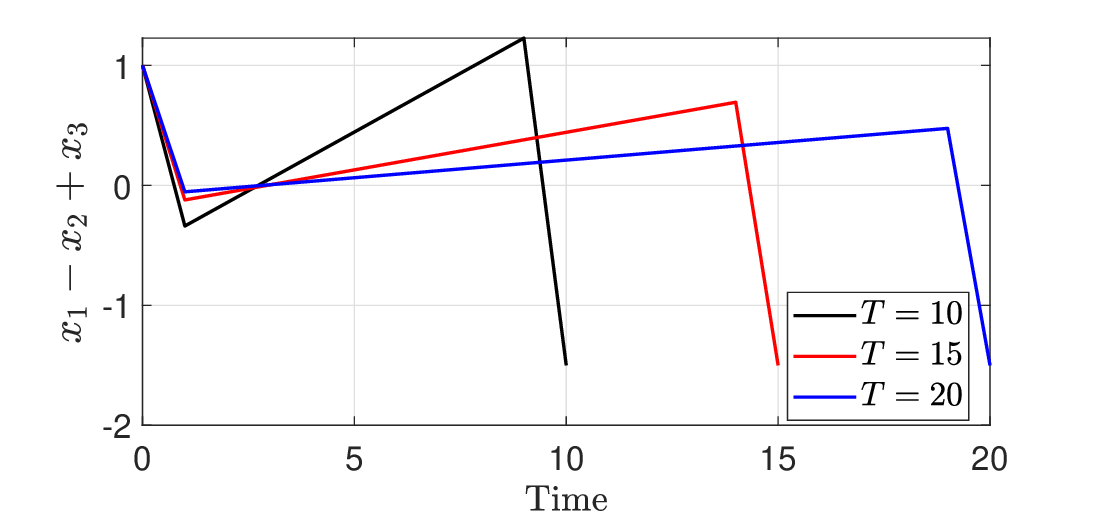}
\caption{Example 1. No subspace turnpike behavior (implicit midpoint rule; left) in contrast to using DDR (right).}
\label{fig:example_1_DDR}
\end{figure}

{\it Example 1:} We consider the %
minimal energy control problem discussed in \cite[Section 5.1]{Schaller21} 
with initial condition $x(0) = [1,\ 1,\ 1]^\top$ and $x(T)=[-1.2,\ -0.7,\ -1]^\top$ and $u_{\operatorname{max}}=50$. 
We obtain the subspace $M = \operatorname{ker}R^{\nicefrac{1}{2}}QJ_-^{-1} = \{ x\in\rline^3\ |\ x_1-x_2+x_3=0\}$. Observe that $\cscr\ne\emptyset$, as required in Theorem \ref{thm:sum_turnpike}. For the DDR case, we also retrieve strict dissipativity w.r.t $\mathcal M$ via Theorem \ref{th:DTOCP_strict_dissip} as Assumption~\ref{asm:manifold} holds with $s=1$, see also \cite[Remark 15]{Karsai2024}. However, it is not possible to verify the assumption of Proposition~\ref{pr:mid_pt_result} apriori to establish the dissipativity property via implicit framework. Infact, in Fig.~\ref{fig:example_1_DDR}, we observe that an implicit midpoint discretization of the optimal control problem does not yield turnpike behavior with respect to the subspace $M$. However, with DDR, the optimal control problem exhibits turnpike behavior as depicted in Fig.~\ref{fig:example_1_DDR} confirming the findings of Theorem~\ref{thm:sum_turnpike} and showing that the assumption $Bu^*_k\in \operatorname{ker}R^{\nicefrac12}QJ_{-}^{-1}$ of Proposition~\ref{pr:mid_pt_result} is not satisfied.\footnote{Indeed, if the implicit midpoint discretization would yield a strictly dissipative optimal control problem, an analogous result as Theorem~\ref{thm:sum_turnpike} would imply a turnpike behavior.}  %

\begin{figure}
\centering   
\includegraphics[width=.45\textwidth]{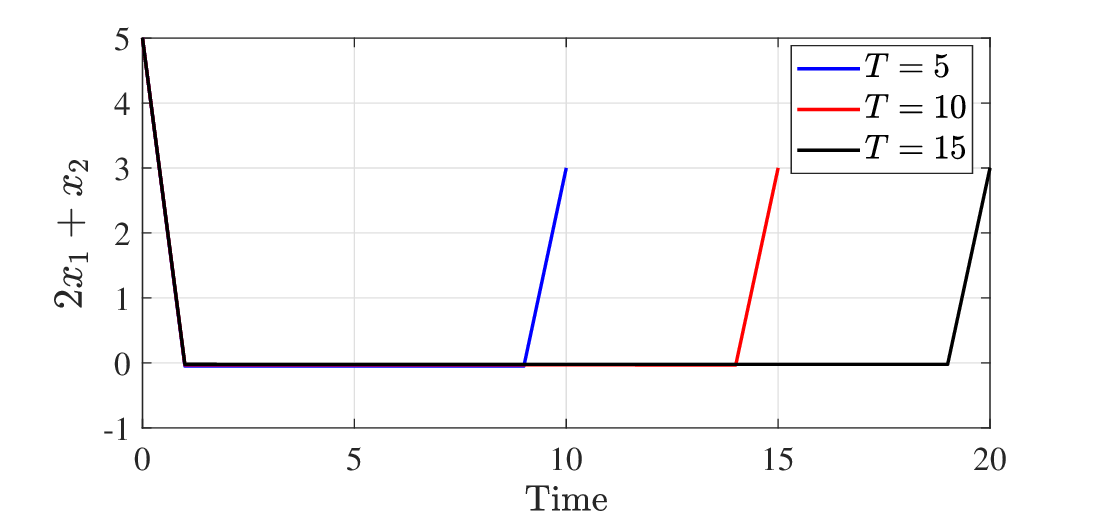}
\caption{Example 2. Turnpike behavior with respect to the manifold $2x_1+x_2=0$.}
\label{fig:example_2}
\end{figure}

{\it Example 2:} We consider the nonlinear pH system described in \cite[Example 17]{Karsai2024}.  
Using Proposition \ref{pr:DDR_quadratic_Ham}, we obtain the DDR with matrices %
$$
\resizebox{.22\textwidth}{!}{$J_-(x_k) = \bbm{1+\tfrac{1}{4}(4\|x_k\|^2+1)^2 & -0.5\\1 & 1}$},\; \resizebox{.22\textwidth}{!}{$J_+(x_k) =\bbm{1-\tfrac{1}{4}(4\|x_k\|^2+1)^2 & 0.5\\-1 & 1}$}.$$
For the DDR considered in this example with $x_k = [x_{1,k} \ x_{2,k}]$, we obtain the function 
\begin{equation}\label{eq:example_subspace}
R^{\tfrac{1}{2}}(x_k)QJ_-^{-1}(x_k)x_k = \frac{h_1(x_k)}{h_2(x_k)}\bbm{2x_{1,k}+x_{2,k} \\ 0 },
\end{equation}
where $h_1(x_k) = 8\|x_k\|^2+2$ and $h_2(x_k) = 16\|x_k\|^4+8\|x_k\|^2+7$. 
For nonlinear pH systems, it is challenging to verify all the conditions in Assumption \ref{asm:manifold}, see also \cite[Remark 16]{Karsai2024}.  However, the real analyticity of the map \(x_k \mapsto R(x_k)^\frac{1}{2}QJ^{-1}_-(x_k)x_k\) and the Assumption \ref{asp:compact_trajectory} yields a similar estimate as \eqref{eq:estimate_manifold} \cite[Remark 6]{Karsai2024}. We consistently observe turnpike behavior for this nonlinear example in numerical simulations. For instance, with initial condition $x_0= [2\ 1]^\top$ and $x_N = [1\ 1]^\top$, we have $\cscr\ne\emptyset$ and corresponding state values exhibit turnpike behavior with respect to the manifold $\mscr = \{x_1\in\rline, x_2\in\rline|2x_1+x_2 = 0\}$ as shown in Fig.~\ref{fig:example_2}.

\section{Conclusion}\label{se:conclusion}
In this paper, we studied the state transition problem for discrete-time nonlinear pH systems with quadratic Hamiltonian, under minimal energy constraint. %
We provided two results ensuring strict dissipativiy of discrete-time energy-optimal control problems, that is, using the implicit midpoint rule and the difference and differential representation.
Under a mild reachability assumption, we also proved that the discrete version of the optimal control problem exhibits turnpike behavior. Finally, we validated our findings using numerical examples. 
This work lays the foundation for future research considering performance and stability analysis of MPC for pH systems.
\bibliographystyle{ieeetr}
\bibliography{refs}

\begin{thebibliography}{10}

\bibitem{JacZ12}
B.~Jacob and H.~Zwart, {\em Linear port-{H}amiltonian systems on
  infinite-dimensional spaces}.
\newblock Operator Theory: Advances and Applications, 223, Basel CH:
  Birkh{\"a}user/Springer Basel AG, 2012.

\bibitem{SchaJelt14}
A.~van~der Schaft and D.~Jeltsema, ``Port-{H}amiltonian systems theory: An
  introductory overview,'' {\em Foundations and Trends{\textregistered} in
  Systems and Control}, vol.~1, no.~2-3, pp.~173--378, 2014.

\bibitem{MehU22}
V.~Mehrmann and B.~Unger, ``Control of port-{H}amiltonian
  differential-algebraic systems and applications,'' {\em Acta Numerica},
  vol.~32, p.~395–515, 2023.

\bibitem{Ortega02}
R.~Ortega, A.~{van der Schaft}, B.~Maschke, and G.~Escobar, ``Interconnection
  and damping assignment passivity-based control of port-controlled
  {H}amiltonian systems,'' {\em Automatica}, vol.~38, no.~4, pp.~585--596,
  2002.

\bibitem{MuelWort18}
M.~A. M{\"u}ller and K.~Worthmann, ``Quadratic costs do not always work in
  {MPC},'' {\em Automatica}, vol.~82, pp.~269--277, 2017.

\bibitem{Sch17}
A.~van~der Schaft, {\em L2-gain and passivity techniques in nonlinear control,
  3rd Edition}.
\newblock Springer, 2017.

\bibitem{SepJanKok12}
R.~Sepulchre, M.~Jankovic, and P.~V. Kokotovic, {\em Constructive nonlinear
  control}.
\newblock Springer Science \& Business Media, 2012.

\bibitem{ScZeWo:24}
M.~Schaller, A.~Zeller, M.~Böhm, O.~Sawodny, C.~Tarín, and K.~Worthmann,
  ``Energy-optimal control of adaptive structures,'' {\em at -
  Automatisierungstechnik}, vol.~72, no.~2, pp.~107--119, 2024.

\bibitem{Schaller21}
M.~Schaller, F.~Philipp, T.~Faulwasser, K.~Worthmann, and B.~Maschke, ``Control
  of port-{H}amiltonian systems with minimal energy supply,'' {\em European
  Journal of Control}, vol.~62, pp.~33--40, 2021.

\bibitem{Faulwasser2022a}
T.~Faulwasser, B.~Maschke, F.~Philipp, M.~Schaller, and K.~Worthmann, ``Optimal
  control of port-{Hamiltonian} descriptor systems with minimal energy
  supply,'' {\em SIAM Journal on Control and Optimization}, vol.~60, no.~4,
  pp.~2132--2158, 2022.

\bibitem{philipp2024optimal}
F.~M. Philipp, M.~Schaller, K.~Worthmann, T.~Faulwasser, and B.~Maschke,
  ``Optimal control of port-{H}amiltonian systems: energy, entropy, and
  exergy,'' {\em Systems \& Control Letters}, vol.~194, 105942, 2024.

\bibitem{ZanoFaul18}
M.~Zanon and T.~Faulwasser, ``Economic {MPC} without terminal constraints:
  Gradient-correcting end penalties enforce asymptotic stability,'' {\em
  Journal of Process Control}, vol.~63, pp.~1--14, 2018.

\bibitem{csen2023stage}
G.~D. {\c{S}}en, M.~Schaller, and K.~Worthmann, ``Stage-cost design for optimal
  and model predictive control of linear port-{Hamiltonian} systems: Energy
  efficiency and robustness,'' {\em PAMM}, vol.~23, no.~4, 202300296, 2023.

\bibitem{kotyczka2019}
P.~Kotyczka and L.~Lefevre, ``Discrete-time port-{H}amiltonian systems: A
  definition based on symplectic integration,'' {\em Systems \& Control
  Letters}, vol.~133, 104530, 2019.

\bibitem{Monaco1998}
S.~Monaco and D.~Normand-Cyrot, ``Discrete-time state representations, a new
  paradigm, perspectives in control: Theory and applications,'' in {\em
  Perspectives in Control: Theory and Applications}, pp.~191--203, Springer
  London, 1998.

\bibitem{Moreschini2019}
A.~Moreschini, M.~Mattioni, S.~Monaco, and D.~Normand-Cyrot, ``Discrete
  port-controlled {Hamiltonian} dynamics and average passivation,'' in {\em
  58th IEEE Conference on Decision and Control}, pp.~1430--1435, 2019.

\bibitem{FaulMull18}
T.~Faulwasser, L.~Gr{\"u}ne, and M.~A. M{\"u}ller, ``Economic nonlinear model
  predictive control,'' {\em Foundations and Trends{\textregistered} in Systems
  and Control}, vol.~5, no.~1, pp.~1--98, 2018.

\bibitem{Karsai2024}
A.~Karsai, ``Manifold turnpikes of nonlinear port-{Hamiltonian} descriptor
  systems under minimal energy supply,'' {\em Mathematics of Control, Signals,
  and Systems}, vol.~36, pp.~707--728, Sep 2024.

\bibitem{AronMehr24}
M.~S. Aronna and V.~Mehrmann, ``Conditions for singular optimal control of
  port-{H}amiltonian systems,'' {\em Preprint arXiv:2407.03213}, 2024.

\bibitem{koelsch2021optimal}
L.~K{\"o}lsch, P.~Jan{\'e}~Soneira, F.~Strehle, and S.~Hohmann, ``Optimal
  control of port-{H}amiltonian systems: A continuous-time learning approach,''
  {\em Automatica}, vol.~129, 109725, 2021.

\bibitem{massaroli2022optimal}
S.~Massaroli, M.~Poli, F.~Califano, J.~Park, A.~Yamashita, and H.~Asama,
  ``Optimal energy shaping via neural approximators,'' {\em SIAM Journal on
  Applied Dynamical Systems}, vol.~21, no.~3, pp.~2126--2147, 2022.

\bibitem{stegink2017port}
T.~Stegink, C.~De~Persis, and A.~van~der Schaft, ``A port-{H}amiltonian
  approach to optimal frequency regulation in power grids,'' {\em IEEE
  Transactions on Automatic Control}, vol.~62, no.~11, pp.~2612--2623, 2017.

\bibitem{gernandt2025port}
H.~Gernandt and M.~Schaller, ``Port-{H}amiltonian structures in
  infinite-dimensional optimal control: Primal--dual gradient method and
  control-by-interconnection,'' {\em Systems \& Control Letters}, vol.~197,
  106030, 2025.

\bibitem{book:Grune2017}
L.~Gr{\"u}ne and J.~Pannek, {\em Nonlinear model predictive control}.
\newblock Springer, 2017.

\bibitem{GrunMull16}
L.~Gr{\"u}ne and M.~A. M{\"u}ller, ``On the relation between strict
  dissipativity and turnpike properties,'' {\em Systems \& Control Letters},
  vol.~90, pp.~45--53, 2016.

\bibitem{FaulFlass22}
T.~Faulwasser, K.~Fla{\ss}kamp, S.~Ober-Bl{\"o}baum, M.~Schaller, and
  K.~Worthmann, ``Manifold turnpikes, trims, and symmetries,'' {\em Mathematics
  of Control, Signals, and Systems}, vol.~34, no.~4, pp.~759--788, 2022.

\bibitem{MehrMora19}
V.~Mehrmann and R.~Morandin, ``Structure-preserving discretization for
  port-{H}amiltonian descriptor systems,'' in {\em 58th IEEE Conference on
  Decision and Control}, pp.~6863--6868, 2019.

\bibitem{XiAnPa:17}
M.~Xia, P.~J. Antsaklis, V.~Gupta, and F.~Zhu, ``Passivity and dissipativity
  analysis of a system and its approximation,'' {\em IEEE Transactions on
  Automatic Control}, vol.~62, no.~2, pp.~620--635, 2017.

\bibitem{MoBiAs:25}
A.~Moreschini, M.~Bin, A.~Astolfi, and T.~Parisini, ``A generalized passivity
  theory over abstract time domains,'' {\em IEEE Transactions on Automatic
  Control}, vol.~70, no.~1, pp.~2--17, 2025.

\bibitem{Moreschini2021Thesis}
A.~Moreschini, {\em Modeling and control of discrete-time and sampled-data
  port-{Hamiltonian} systems}.
\newblock PhD thesis, Universit{\`a} degli Studi di Roma La Sapienza, 2021.
\newblock \url{https://theses.hal.science/tel-03628359v1}.

\bibitem{MoMoNo:19}
A.~Moreschini, S.~Monaco, and D.~Normand-Cyrot, ``{Gradient and Hamiltonian
  dynamics under sampling},'' {\em IFAC-PapersOnLine}, vol.~52, no.~16,
  pp.~472--477, 2019.

\bibitem{MoNoTi:11}
S.~Monaco, D.~Normand-Cyrot, and F.~Tiefensee, ``{Sampled-data stabilization; A
  PBC approach},'' {\em IEEE Transactions on Automatic Control}, vol.~56,
  no.~4, pp.~907--912, 2011.

\end{thebibliography}
\end{document}